\documentclass[a4paper,reprint,12pt,aps,tightenlines,onecolumn,notitlepage,longbibliography]{revtex4-1}
\usepackage{placeins}
\usepackage{amsmath}
\usepackage{amssymb}
\usepackage{cancel}
\usepackage{stmaryrd}
\usepackage{amsfonts}
\usepackage[T1]{fontenc}
\usepackage{bm}
\usepackage{color}
\usepackage{graphicx}
\DeclareSymbolFont{operators}{OT1}{cmr}{m}{n}
\DeclareSymbolFont{letters}{OML}{cmm}{m}{it}
\DeclareSymbolFont{symbols}{OMS}{cmsy}{m}{n}
\DeclareSymbolFont{largesymbols}{OMX}{cmex}{m}{n}
\usepackage[hyperindex,breaklinks]{hyperref}
\hypersetup{
    pdfborder={0 0 0},
    allbordercolors={0 0 0},
    unicode=false,          
    pdftoolbar=true,        
    pdfmenubar=true,        
    pdffitwindow=false,     
    pdfstartview={FitH},    
    pdfcreator={Eero Hirvijoki},
    pdfproducer={pdflatex},
    pdfnewwindow=true,
    colorlinks=true,
    linkcolor=blue,
    citecolor=blue,
    filecolor=blue,
    urlcolor=blue,  
}
\usepackage{tikz}

\makeatother

\begin{document}

\title{Eliminating poor statistics in Monte-Carlo simulations of fast-ion losses to plasma-facing components and detectors}

\author{Eero Hirvijoki}
\affiliation{Aalto University, Department of Applied Physics, Finland}
\email{eero.hirvijoki@gmail.com}

\date{\today}

\begin{abstract}
With Wendelstein 7-X now up and running, and the construction of ITER proceeding, predicting fast-ion losses to sensitive plasma-facing components and detectors is gaining significant interest. A common recipe to perform such studies is to push a large population of marker particles along their equations of motion, the trajectories randomized with Monte Carlo operators accounting for Coulomb collisions, and to record possible intersections of the marker trajectories with synthetic detectors or areas of interest in the first wall. While straightforward to implement and easy to parallelize, this Forward Monte Carlo (FMC) approach tends to suffer from poor statistics and error estimation as the detector domain is often small: it is difficult to guess how to set up the initial weights and locations of the markers for them to remain representative of the source distribution, yet record enough hits to the detector for good statistics. As an alternative, the FMC method can be replaced with a so-called Backward Monte Carlo (BMC) algorithm. Instead of starting with a given initial marker population, one starts from the end condition at the detector and records how the hit probability evolves backwards in time. The scheme eliminates the statistics issue present in the FMC scheme and may provide more accurate and efficient simulations of fast-ion loss signals. The purpose of this paper is to explain the BMC recipe in the fast-ion setting and to discuss the associated nuances, especially how to negate artificial diffusion. For illustration purposes, our numerical example considers a 1-D stochastic harmonic oscillator as a mock-up of a charged particle.
\end{abstract}


\maketitle
 
\textit{Introduction} -- 
To get a feeling for Backward Monte Carlo (BMC) modeling, we start with a one-dimensional random process $Z_t\in\{z_i\}_{i=1}^N$ that has discrete outcomes depending only on the previous value of the process. We shall assume the process $Z_t$ to satisfy the following "go left -- stay put -- go right" jump probabilities on the $z$-axis
\begin{align}
    \mathbb{E}[Z_{t+\Delta t}=z_{i-1}|Z_t=z_i]&= 
    \begin{cases}
    0,   & i=1\\
    1/3, & 1<i<N\\
    1/2, & i=N
    \end{cases}
    \\
    \mathbb{E}[Z_{t+\Delta t}=z_{i}|Z_t=z_i]&= 
    \begin{cases}
    1/2,   & i=1\\
    1/3, & 1<i<N\\
    1/2, & i=N
    \end{cases}
    \\
    \mathbb{E}[Z_{t+\Delta t}=z_{i+1}|Z_t=z_i]&= 
    \begin{cases}
    1/2,   & i=1\\
    1/3, & 1<i<N\\
    0, & i=N.
    \end{cases}
\end{align}
Then we ask: Given any $Z_0=x\in\{z_i\}_{i=1}^N$, at what probability does $Z_{M\Delta t}=y\in\Omega$ happen? In mathematical terms, this is equal to computing the expectation $\mathbb{E}[Z_{M\Delta t}=y|Z_0=x]$. The Forward Monte Carlo (FMC) recipe to solve this problem involves populating each $z_i$ with multiple marker processes, simulating each marker according to the jump probabilities given above, and, after $M$ time steps, checking whether the marker has the value $y$. The FMC algorithm is straightforward to implement on a computer and trivial to parallelize as each marker process is independent of each other. The method might, however, become computationally expensive if a strict error estimate is needed -- the error is proportional to $1/\sqrt{N_i}$ with $N_i$ the number of markers launched from site $z_i$ -- or if the set $\{z_i\}_{i=1}^N$ is very large.

There is, of course, a simpler solution to the problem. Let us define the function
\begin{align}
    \Phi(z_i,m)=\mathbb{E}[Z_{M\Delta t}=y|Z_{m\Delta t}=z_i],
\end{align}
and note that $\Phi(x,0)=\mathbb{E}[Z_{M\Delta t}=y|Z_0=x]$ would give us the expectation value we are chasing and that $\Phi(x,M)=\mathbf{1}_y(x)$ is the indicator function. Using conditional probabilities, we compute a recursive expression for $\Phi(x,m)$, namely
\begin{align}\label{eq:discrete_iteration}
    \Phi(x,m-1)&=\mathbb{E}[Z_{M\Delta t}=y|Z_{(m-1)\Delta t}=x]\nonumber\\
    &=\sum_{z_i\in\Omega}\mathbb{E}[Z_{M\Delta t}=y|Z_{m\Delta t}=z_i]\mathbb{E}[Z_{m\Delta t}=z_i|Z_{(m-1)\Delta t}=x]\nonumber\\
    &=\sum_{z_i\in\Omega}\Phi(z_i,m)\mathbb{E}[Z_{m\Delta t}=z_i|Z_{(m-1)\Delta t}=x]\nonumber\\
    &\equiv\mathbb{E}[\Phi(Z_{m\Delta t},m)|Z_{(m-1)\Delta t}=x],
\end{align}
valid for any $0<m\leq M$. Updating the function $\Phi(x,m)$, by letting $m$ run from $M$ down to $1$, constitutes to computing how the probability to reach the end condition branches out "backwards" in time or, put more precisely, how the probability depends on the time given for the markers to reach the end condition when this time gradually increases. For example, let us choose $M=3$, an end condition $Z_{3\Delta t}=2$, and the set $\{z_i\}_{i=1}^4=\{1,2,3,4\}$. The resulting branching is illustrated below, with bolded numbers denoting the probability (or $\Phi(z,m)$) of each state to reach the end condition, during each of the three rounds of iterations.
\begin{center}
\begin{tikzpicture}[thick,scale=0.85, every node/.style={transform shape}]
\draw[thick,->] (0,0) -- (4.5,0) node[anchor=north west] {$z$};
\draw[thick,->] (0,0) -- (0,4.5) node[anchor=north west] {$t$};

\foreach \y [evaluate=\y as \yeval using \y-1] in {1,...,4} 
    \draw (1pt,\y cm) -- (-1pt,\y cm) node[anchor=east]
    {$\pgfmathtruncatemacro{\yet}{\yeval}\yet \Delta t$}; 

\foreach \x in {1,2,3,4}
    \draw (\x cm,1pt) -- (\x cm,-1pt) node[anchor=north] {$ \x $};

\node[] at (1,4) {$0$};
\node[] at (2,4) {$1$};
\node[] at (3,4) {$0$};
\node[] at (4,4) {$0$};

\node[] at (1,3) {$\bf{\tfrac{1}{2}}$};
\node[] at (2,3) {$\bf{\tfrac{1}{3}}$};
\node[] at (3,3) {$\bf{\tfrac{1}{3}}$};
\node[] at (4,3) {$\bf{0}$};



\end{tikzpicture}
\begin{tikzpicture}[thick,scale=0.85, every node/.style={transform shape}]
\draw[thick,->] (0,0) -- (4.5,0) node[anchor=north west] {$z$};
\draw[thick,->] (0,0) -- (0,4.5) node[anchor=north west] {$t$};

\foreach \y [evaluate=\y as \yeval using \y-1] in {1,...,4} 
    \draw (1pt,\y cm) -- (-1pt,\y cm) node[anchor=east]
    {$\pgfmathtruncatemacro{\yet}{\yeval}\yet \Delta t$}; 

\foreach \x in {1,2,3,4}
    \draw (\x cm,1pt) -- (\x cm,-1pt) node[anchor=north] {$ \x $};

\node[] at (1,4) {$0$};
\node[] at (2,4) {$1$};
\node[] at (3,4) {$0$};
\node[] at (4,4) {$0$};

\node[] at (1,3) {$\tfrac{1}{2}$};
\node[] at (2,3) {$\tfrac{1}{3}$};
\node[] at (3,3) {$\tfrac{1}{3}$};
\node[] at (4,3) {$0$};

\node[] at (1,2) {$\bf{\tfrac{5}{12}}$};
\node[] at (2,2) {$\bf{\tfrac{7}{18}}$};
\node[] at (3,2) {$\bf{\tfrac{2}{9}}$};
\node[] at (4,2) {$\bf{\tfrac{1}{6}}$};


\end{tikzpicture}
\begin{tikzpicture}[thick,scale=0.85, every node/.style={transform shape}]
\draw[thick,->] (0,0) -- (4.5,0) node[anchor=north west] {$z$};
\draw[thick,->] (0,0) -- (0,4.5) node[anchor=north west] {$t$};

\foreach \y [evaluate=\y as \yeval using \y-1] in {1,...,4} 
    \draw (1pt,\y cm) -- (-1pt,\y cm) node[anchor=east]
    {$\pgfmathtruncatemacro{\yet}{\yeval}\yet \Delta t$}; 

\foreach \x in {1,2,3,4}
    \draw (\x cm,1pt) -- (\x cm,-1pt) node[anchor=north] {$ \x $};

\node[] at (1,4) {$0$};
\node[] at (2,4) {$1$};
\node[] at (3,4) {$0$};
\node[] at (4,4) {$0$};

\node[] at (1,3) {$\tfrac{1}{2}$};
\node[] at (2,3) {$\tfrac{1}{3}$};
\node[] at (3,3) {$\tfrac{1}{3}$};
\node[] at (4,3) {$0$};

\node[] at (1,2) {$\tfrac{5}{12}$};
\node[] at (2,2) {$\tfrac{7}{18}$};
\node[] at (3,2) {$\tfrac{2}{9}$};
\node[] at (4,2) {$\tfrac{1}{6}$};

\node[] at (1,1) {$\bf{\tfrac{29}{72}}$};
\node[] at (2,1) {$\bf{\tfrac{37}{108}}$};
\node[] at (3,1) {$\bf{\tfrac{7}{27}}$};
\node[] at (4,1) {$\bf{\tfrac{7}{36}}$};

\end{tikzpicture}
\end{center}

Obviously, there is no uncertainty in determining $\Phi(x,m)$ but, where the FMC method is trivially parallel, the BMC algorithm does require a gather operation of the information $\Phi(z_i,m)$ after each time step. Fortunately, for physical processes, the jump probability $\mathbb{E}[Z_{t+\Delta t}=z_j|Z_t=z_i]$ is often peaked and effectively limited to a narrow range. Hence, even if the set $\{z_i\}_{i=1}^N$ grew significantly, domain decomposition methods would likely be effective. 

Before moving on, we make an additional note. In the example described above, we have used a jump probability that depends only on the previous value of the process. In the iterative expression \eqref{eq:discrete_iteration}, the expectation $\mathbb{E}[Z_{m\Delta t}=y|Z_{(m-1)\Delta t}=x]$ could, however, also depend on the time index $m$, accounting, e.g., for temporal changes in conditions the jump probabilities depend on. With this introduction to BMC calculation, we shall apply the principle to model propabilities of fast ions in tokamak or stellarator plasmas to reach preset phase-space domains such as specific wall tiles or particle detectors.

\textit{Traditional modeling of fast-ion-loss signals:} --
Arguably, the Vlasov equation, supplemented with the Landau collision operator, is a sufficient model to describe the distribution function of charged particles in many magnetized plasmas, such as those in tokamaks and stellarators. While the Landau operator is non-linear, in case of fast ions that form a minority population in typical plasmas, the so-called field-particle distribution in the collision operator can be approximated to be a Maxwellian and the self-collisions neglected. The kinetic equation for fast ions can then be assumed linear. 

The resulting, so-called test-particle kinetic equation is a Fokker-Planck equation with a well-known connection to continuous stochastic processes. In a given background plasma, an individual marker particle with coordinates $\bm{Z}_t=(\bm{X}_t,\bm{V}_t)$, sampled from the distribution function of species $\alpha$, is subject to the following stochastic differential equation (SDE) of It\^o kind
\begin{align}
    d\bm{X}_t&=U_{\alpha}^{\bm{x}}(\bm{Z}_t,t)dt,\\
    d\bm{V}_t&=U_{\alpha}^{\bm{v}}(\bm{Z}_t,t)dt+\bm{\mu}_{\alpha}(\bm{Z}_t,t)dt+\bm{\sigma}_{\alpha}(\bm{Z}_t,t)\cdot d\bm{W}_t.
\end{align}
The vector field $U_{\alpha}(\bm{z},t)=(U_{\alpha}^{\bm{x}}(\bm{z},t),U_{\alpha}^{\bm{v}}(\bm{z},t))$ has the components
\begin{align}
    U_{\alpha}^{\bm{x}}(\bm{z},t)&=\bm{v}, \\ U_{\alpha}^{\bm{v}}(\bm{z},t)&=e_{\alpha}(\bm{E}(\bm{x},t)+\bm{v}\times\bm{B}(\bm{x},t))/m_{\alpha},
\end{align}
with $\bm{E}$ and $\bm{B}$ the electric and magnetic fields and $e_{\alpha}$ and $m_{\alpha}$ the charge and mass of species $\alpha$, and represents time-dependent Hamiltonian flow
\begin{align}
    \bm{z}(t)\mapsto\varphi_{U_{\alpha}}^{s,t}(\bm{z}(t))&=\bm{z}(t+s),\\
    \frac{d\varphi^{s,t}_{U_{\alpha}}(\bm{z})}{ds}\Big\vert_{s=0}&=U_{\alpha}(\bm{z},t).
\end{align}
The vector $\bm{\mu}_{\alpha}$ and the matrix $\bm{\sigma}_{\alpha}$ appearing in the SDE for $\bm{Z}_t$ originate from the test-particle collision operator, with Cartesian components given by
\begin{align}
    \mu^i_{\alpha}(\bm{x},\bm{v},t)&=\nu_{\alpha}(\bm{x},\bm{v},t) v^i,\\
    \sigma^{ij}_{\alpha}(\bm{x},\bm{v},t)&=\sqrt{2D_{\alpha\parallel}(\bm{x},\bm{v},t)}\frac{v^iv^j}{|\bm{v}|^2}+\sqrt{2D_{\alpha\perp}(\bm{x},\bm{v},t)}\left(\delta^{ij}-\frac{v^iv^j}{|\bm{v}|^2}\right),
\end{align}
and the coefficients $\nu_{\alpha}$, $D_{\alpha\parallel}$, and $D_{\alpha\perp}$ being 
\begin{align}
    \nu_{\alpha}&=\sum_{\beta}\frac{n_{\beta}c_{\alpha\beta}}{m^2_{\alpha}}\left(1+\frac{m_{\alpha}}{m_{\beta}}\right)\left(\frac{m_{\beta}}{2T_{\beta}}\right)^{3/2}\frac{1}{y}\frac{d}{dy}\left(\frac{d^2\Psi(y)}{dy^2}+\frac{2}{y}\frac{d\Psi(y)}{dy}\right)\Big\vert_{y=y_{\beta}},\\
    D_{\alpha\parallel}&=\sum_{\beta}\frac{n_{\beta} c_{\alpha\beta}}{m^2_{\alpha}}\sqrt{\frac{m_{\beta}}{2T_{\beta}}}\frac{d^2\Psi(y)}{dy^2}\Big\vert_{y=y_{\beta}},\\
    D_{\alpha\perp}&=\sum_{\beta}\frac{n_{\beta} c_{\alpha\beta}}{m^2_{\alpha}}\sqrt{\frac{m_{\beta}}{2T_{\beta}}}\frac{1}{y}\frac{d \Psi(y)}{dy}\Big\vert_{y=y_{\beta}}.
\end{align}
The argument $y_{\beta}$, at which the function $\Psi(y)=\pi^{-1/2}\exp(-y^2)+(y+1/(2y))\text{erf}(y)$ and its derivatives are to be evaluated, is the particle speed normalized to the thermal velocity, i.e.,  $y_{\beta}=\sqrt{m_{\beta}|\bm{v}|^2/2T_{\beta}}$, with $n_{\beta}(\bm{x},t)$ and $T_{\beta}(\bm{x},t)$ the density and temperature of species $\beta$. The coefficient $c_{\alpha\beta}=e_{\alpha}^2e_{\beta}^2\ln\Lambda_{\alpha\beta}/(8\pi \varepsilon_0^2)$ involves the Coulomb logarithm $\ln\Lambda_{\alpha\beta}$ and, finally, $\bm{W}_t\sim\mathcal{N}(\bm{0},t\mathbf{I})$ is the standard multivariate Wiener process. In essence, $\nu_{\alpha}$ acts as a friction coefficient, slowing the particle down, whereas $D_{\alpha\parallel}$ results in velocity diffusion along the particle velocity and $D_{\alpha\perp}$ in the direction perpendicular to it.

The above set of equations and expressions describes how a test particle with phase-space coordinates $\bm{Z}_t=(\bm{X}_t,\bm{V}_t)$ behaves in a plasma, including the fast Larmor gyration around the magnetic field line. One could write an analogous set of equations describing the guiding-center motion of test particles and, also, further approximate the expressions for $\bm{\mu}$ and $\bm{\sigma}$ to reflect the speed of the fast ions with respect to a thermal plasma.

Out in the field, deep in the trenches, several numerical codes \cite{SPIRAL_reference,ORBIT_reference,LOCUST_reference,ASCOT4_reference,VENUS-LEVIS_reference,OFMC_reference,RIO_reference,FOCUS_reference,KORC_reference,GYCAVA_reference} use the equations above -- or their guiding-center counterparts -- to perform FMC simulations of various fast ion (or electron) signals such as the energy flux onto certain wall tiles or the distribution of fast ions recorded by a fast-ion-loss detector (FILD). In practice, each code implements the following algorithm: (i) Assign a marker particle with initial coordinates $\bm{Z}_0=(\bm{X}_0,\bm{V}_0)$. (ii) Iterate the phase-space position of the marker by advancing the Hamiltonian flow along the vector field $U_{\alpha}$ with the map
\begin{align}\label{eq:hamiltonian_step}
    \bm{Z}\mapsto\varphi_{U_{\alpha}}^{\Delta t,t}(\bm{Z}),
\end{align}
and by updating the marker velocity to account for Coulomb collisions with the Euler-Maryama scheme
\begin{align}\label{eq:coulomb_step}
(\bm{X},\bm{V})\mapsto\varphi_C^{\Delta t,t}(\bm{X},\bm{V})=(\bm{X},\bm{V}+\bm{\mu}\left(\bm{Z},t\right)\Delta t+\bm{\sigma}\left(\bm{Z},t\right)\cdot\bm{\xi}\sqrt{\Delta t}),
\end{align}
where $\bm{\xi}\sim\mathcal{N}(\bm{0},\mathbf{I})$ are random numbers obeying standard multivariate normal distribution. (iii) If the trajectory intersects a wall tile or a diagnostic, represented by a phase-space domain $\Omega$, record a hit, accumulate the recorded quantity, and terminate the marker if necessary. In this process, a large number of marker particles is used to populate the phase-space, and often only a few hit the detector, resulting in poor statistics.

{\it BMC approach:} -- Fast-ion signals could be equivalently computed from the expression
\begin{align}\label{eq:signal}
    I(\Omega,t)=\int S(\bm{z}')\mathbb{E}[P(\bm{Z}_{\tau};\Omega,t)|\bm{Z}_0=\bm{z}']d\bm{z}'.
\end{align}
Here $I$ is the signal strength, $\Omega$ is the domain from which the signal is produced, and $t$ is the time interval over which the marker particles are allowed to evolve from their initial state and possibly intersect the domain $\Omega$. The signal is then computed as a sum of the source particles, represented by $S(\bm{z})$, weighted with the probability $\mathbb{E}[P(\bm{Z}_{\tau};\Omega,t)|\bm{Z}_0=\bm{z}]$ for a marker from $\bm{z}$ to reach the domain $\Omega$ during the interval $[0,t]$. The expression $P(\bm{Z}_{\tau};\Omega,t)$ is a statement of the end condition, and could read "a marker $\bm{Z}_{\tau}$ has intersected the domain $\Omega$ with $\tau\in[0,t]$". Obviously, there is no Monte Carlo sampling error in the signal $I$ unless evaluation of the expectation $\mathbb{E}[P(\bm{Z}_{\tau};\Omega,t)|\bm{Z}_0=\bm{z}]$ introduces such. 

As we saw in the introductory part, an expectation value to reach an end condition can be computed in a deterministic fashion using the BMC algorithm: it was demonstrated explicitly for a process with discrete outcomes. For a process with continuous outcomes, such as the trajectory of fast ions resulting from the equations \eqref{eq:hamiltonian_step} and \eqref{eq:coulomb_step}, the approach is almost exactly the same. The only significant difference is that, in computing the expectation value, the sum over different outcomes has to be computed as an integral over the associated probability measure. In our case this means that each realization of the fast ion trajectory $\bm{Z}_{m\Delta t}$ should be interpreted as a function of the random variable $\bm{\xi}\sim\mathcal{N}(0,\mathbf{I})$ and the expectation ultimately taken with respect to the probability density of $\bm{\xi}$. 

If we now define the function
\begin{align}
    \Phi(\bm{z},m)=\mathbb{E}[P(\bm{Z}_{\tau};\Omega,M\Delta t)|\bm{Z}_{m\Delta t}=\bm{z}], \qquad \tau\in[m\Delta t,M\Delta t],
\end{align}
where $P(\bm{Z}_{\tau};\Omega,M\Delta t)$ refers to the end condition and $m$ labels the time step index, we may yet again build upon the conditionality of the probabilities and compute a recursion formula similar to that argued in the introduction:
\begin{align}\label{eq:recursive_plain}
    \Phi(\bm{z},m-1)&=\mathbb{E}[P(\bm{Z}_{\tau};\Omega,M\Delta t)|\bm{Z}_{(m-1)\Delta t}=\bm{z}]\nonumber\\
    &=\int\mathbb{E}[P(\bm{Z}_{\tau};\Omega,M\Delta t)|\bm{Z}_{m\Delta t}=\bm{z}']\mathbb{E}[\bm{Z}_{m\Delta t}=\bm{z}'|\bm{Z}_{(m-1)\Delta t}=\bm{z}]d\bm{z}'\nonumber\\
    &=\int \Phi(\bm{z}',m)\mathbb{E}[\bm{Z}_{m\Delta t}=\bm{z}'|\bm{Z}_{(m-1)\Delta t}=\bm{z}]d\bm{z}'\nonumber\\
    &\equiv\mathbb{E}[\Phi(\bm{Z}_{m\Delta t},m)|\bm{Z}_{(m-1)\Delta t}=\bm{z}].
\end{align}
To evaluate the expectation value, we rely on the fact that each realization of the fast-ion trajectory from point $\bm{Z}_{(m-1)\Delta t}$ to $\bm{Z}_{m\Delta t}$, computed as a composition of the maps \eqref{eq:hamiltonian_step} and \eqref{eq:coulomb_step}, is a time-dependent random map which we denote by
\begin{align}\label{eq:random_map}
    \bm{z}\mapsto\varphi^{\Delta t,(m-1)\Delta t}(\bm{z},\bm{\xi}).
\end{align}
Taking this into account, the recursive formula \eqref{eq:recursive_plain} can be expressed as
\begin{align}\label{eq:recursive}
    \Phi(\bm{z},m-1)=\int_{\mathbb{R}^3}\Phi(\varphi^{\Delta t,(m-1)\Delta t}(\bm{z},\bm{\xi}),m)\frac{\exp[-\bm{\xi}^2/2]}{(2\pi)^{3/2}}d\bm{\xi}.
\end{align}
In computing fast ion signals, one may thus use this recursive integral expression to evaluate the necessary expectation value, without introducing Monte Carlo sampling noise. As of yet, the only source of error in the formula \eqref{eq:recursive} results from the integration of the stochastic differential equations to construct the fast-ion trajectory. As the map $\bm{z}\mapsto\varphi^{\Delta t,(m-1)\Delta t}(\bm{z},\bm{\xi})$ is shared with the FMC scheme, the BMC expression \eqref{eq:recursive} as such is no less accurate than the process of following markers in the FMC scheme.

Although the BMC formula might not immediately ring a bell in the fast-ion community, it has been put to good use elsewhere in the past. The methodology has helped to estimate, e.g., runaway-electron generation rates in a tokamak \cite{Zhang-DiegoPoP2017-backwards-monte-carlo}, to price options in finance \cite{bormetti_callegaro_livieri_pallavicini_2018}, and even to craft a Langevin approach to multi-scale modeling \cite{hirvijoki_langevin:2018PhPl} and a fluid-kinetic framework for self-consistent runaway-electron modeling \cite{hirvijoki_fluid_kinetic:2018PhPl}. Furthermore, the BMC algorithm shares a close ideology with the adjoint methods for Fokker-Planck equations which have been used to study, e.g., current drive \cite{karney_current_1986} and runaway electrons \cite{Liu_2016_adjoint}. 



\textit{An example:} -- We will demonstrate the numerical implementation of the expression \eqref{eq:recursive} using a 1-D stochastic harmonic oscillator as a mock-up as it displays the most essential characteristics of charged-particle motion in a plasma: the circular Hamiltonian motion is akin to the orbit topologies of charged particles with their gyro, bounce, and passing motion, and the stochastic addition to the velocity mimics the effects of Coulomb collisions, resulting in transport across the isosurfaces of energy. More importantly, this simplified model allows us to illustrate the effect of artificial diffusion which might arise if the marker evolution is dominated by Hamiltonian dynamics -- this is the case for fast ions in tokamak or stellarator plasmas -- and proper care is not taken while advancing the marker trajectories. 

Our model stochastic differential equations read
\begin{align}
    dX_t&=V_tdt,\\
    dV_t&=-X_tdt+\sigma dW_t,
\end{align}
and, with $\sigma$ set to zero, the marker characteristics would draw circles centered around the point $(x,v)=(0,0)$. With $\sigma$ being non-zero, we expect deviations from the circles and, were we to draw a rectangular box centered at $(x,v)=(0,0)$, eventual intersections of the marker trajectories with the box. To construct the map for moving markers around, we use the symplectic Euler for the Hamiltonian part and the Euler-Maryama scheme for the stochastic part, providing us with the explicit maps
\begin{align}
    \label{eq:symplectic_euler}
    (x,v)\mapsto\varphi^{\Delta t}_H(x,v)&=(x(1-\Delta t^2)+\Delta t v,v-\Delta t x)
    \\
    \label{eq:euler_maryama}
    (x,v)\mapsto\varphi^{\Delta t}_S(x,v)&=(x,v+\sigma\sqrt{\Delta t}\xi).
\end{align}
We will then combine these maps into the following composition
\begin{align}
    (x,v)\mapsto\varphi^{\Delta t}_S\circ\underbrace{\varphi_H^{\Delta t/n}\circ\dots\circ\varphi_H^{\Delta t/n}(x,v)}_{n}\equiv\varphi^{\Delta t}(x,v,\xi),
\end{align}
where the number of symplectic Euler substeps, $n$, can be chosen freely. The actual time step $\Delta t$ will be chosen so that the condition $\sigma^2\Delta t \ll 1$ is satisfied.

The reasoning behind multiple, optional Hamiltonian substeps is to minimize artificial numerical diffusion when $\sigma$ is small but yet finite: In computing the recursive update \eqref{eq:recursive}, the function $\Phi$ has to be presented in some finite form, typically in a mesh. As the map $\bm{z}\mapsto\varphi^{\Delta t,(m-1)\Delta t}(\bm{z},\bm{\xi})$ will rarely result in locations that coincide with the mesh points, interpolation of $\Phi$ will be needed. This interpolation will introduce artificial diffusion in a similar way as is encountered in semi-Lagrangian schemes for advection equations. By choosing the time step $\Delta t$ to allow for maximum effect of physical diffusion to take place, the effect of artificial diffusion can be negated. If $\sigma$ is small, a large time step can be allowed, and hence sub steps become necessary for correctly evaluating the Hamiltonian contribution.

In our demonstration simulation, we choose a rectangular box for the computational domain, with corners at $(x,v)=(-2,-2),(2,-2),(2,2),(-2,2)$. The boundaries of the box will be indicated with a purple contour. The vertical lines at $x=\pm 2$ can be seen to represent the wall surface of a tokamak or a stellarator and, to model a FILD, we consider segments of the vertical lines, with each segment naturally corresponding to a different velocity regime, mimicking the different energy channels of a real FILD. The function $\Phi(x,v)$ is represented on a fixed 2-D rectilinear mesh, with 51 mesh points in both $x$ and $v$ directions, and linear interpolation applied in between the node points. The integral in \eqref{eq:recursive}, in our mock-up case a 1-D Gaussian integral, is computed with a Gauss-Hermite quadrature. The characteristic time scale in the problem, the Hamiltonian orbit time, is $2\pi$ for all closed Hamiltonian orbits. 

\begin{figure}[!h]
    \includegraphics[width=.32\linewidth]{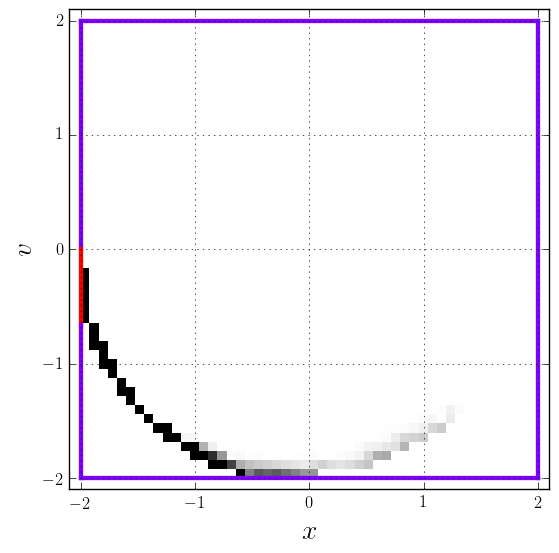}
    \includegraphics[width=.32\linewidth]{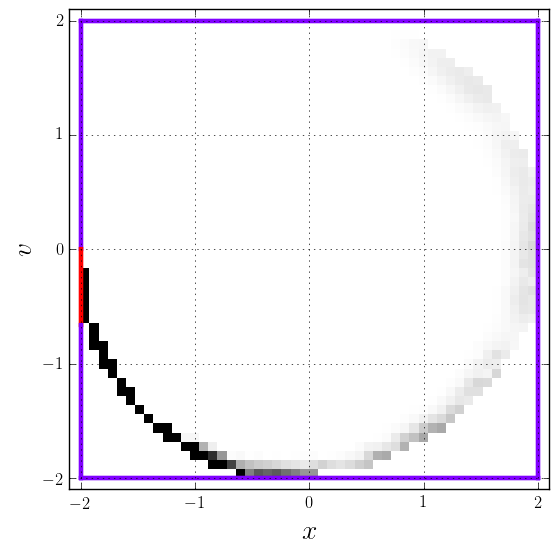}
    \includegraphics[width=.32\linewidth]{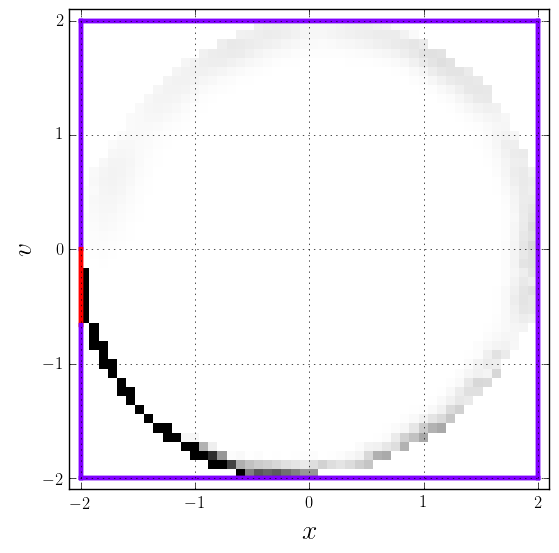}
    \includegraphics[width=.32\linewidth]{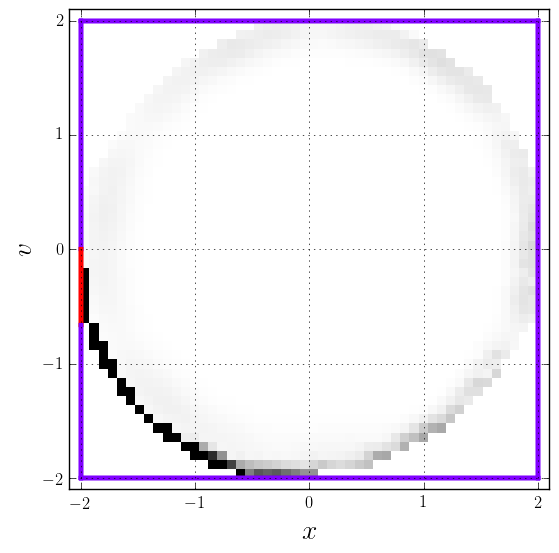}
    \includegraphics[width=.32\linewidth]{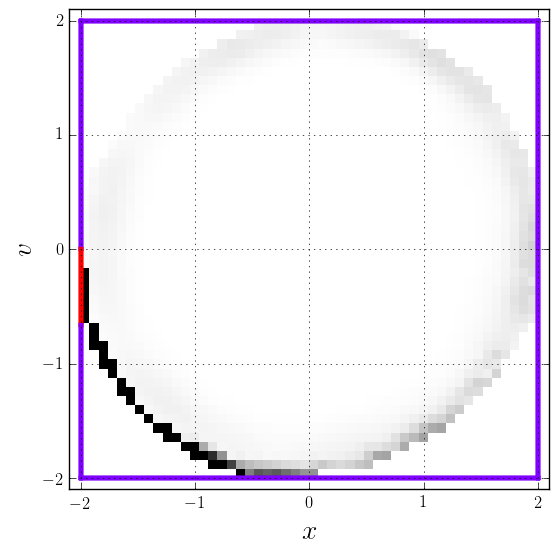}
    \includegraphics[width=.32\linewidth]{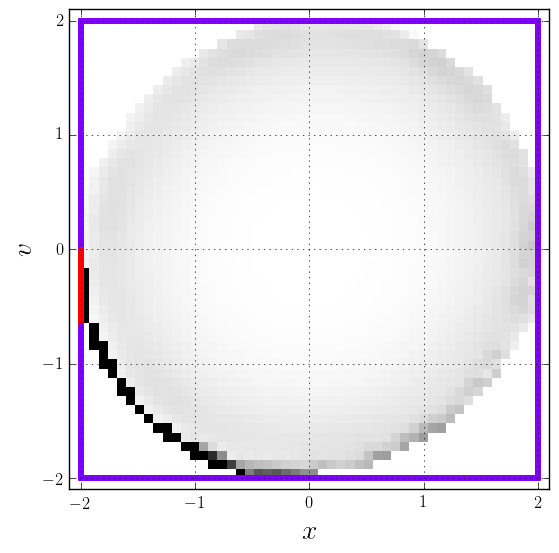}
    \caption{Probability $\Phi(x,v)$ for a marker to reach the red segment in a given time. Chosen parameters were $\sigma=0.1$, $\Delta t=0.01/\sigma^2=1.0$, $n=12$. The evolution is from left to right and top to bottom, with times of the frames corresponding to $t=(2,4,6,8,10,20\pi)$. Black color refers to one, white to zero, and gray in between.}
    \label{fig:time_evolution}
\end{figure}

\begin{figure}[!h]
    \includegraphics[width=.32\linewidth]{t_10.png}
    \includegraphics[width=.32\linewidth]{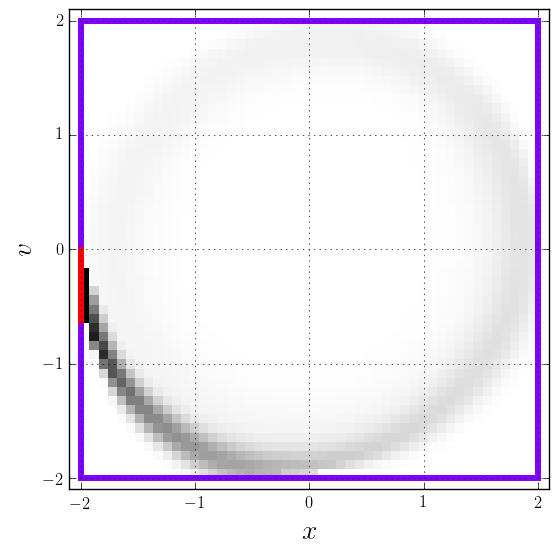}
    \caption{Illustration of artificial diffusion. On the left, the $t=10$ frame of Fig. \ref{fig:time_evolution} repeated. On the right, the same simulation with a shorter time step of $\Delta t=0.1$. Artificial diffusion akin to what is present in semi-Lagrangian methods for advection equations dominates over the physical diffusion.}
    \label{fig:artificial_diffusion_comparison}
\end{figure}

In figure \ref{fig:time_evolution}, we see how $\Phi(x,v)$, the probability for markers to reach the red segment in given time, branches out backwards in time, reaching further out with more time given for markers to reach the red segment. One should note that the domain of the red segment would be entirely inaccessible for closed Hamiltonian orbits without diffusion. The example thus demonstrates that the BMC algorithm is able to capture probabilities of particles to reach domains that have a very narrow window of entrance, such as the FILD in a tokamak or stellarator environment. For parameters in this simulation, we chose $\sigma=0.1$, $\Delta t=0.01/\sigma^2=1.0$, and $n=12$. 
Figure \ref{fig:artificial_diffusion_comparison}, on the other hand, illustrates the effect of choosing the time step for the recursive iteration correctly. Too short a time step results in artificial diffusion similarly as in semi-Lagrangian schemes for advection equations and is clearly visible in the right frame of the figure, in the form of extra smearing of the function~$\Phi(x,v)$.

\textit{Summary and concluding thoughts} -- In this paper, we have depicted a scheme to eliminate statistical errors from Monte Carlo simulations of fast-ion-loss signals. This is a problem that plagues, e.g., estimations of FILD signals in tokamak and stellarator plasmas as very few of the vast amount of marker particles used in the forward Monte Carlo simulations reach the final target domain. Our results with the mock-up, the 1-D stochastic harmonic oscillator, are encouraging in that the scheme is practical and easy to program but also reveal that further studies should perhaps be conducted to experiment with mesh generation techniques before plunging into the deep end of the pool, the fast ions in tokamak and stellarator geometry. One can, e.g., interpret from the figures that a rectilinear mesh might not offer the best fit to circular Hamiltonian orbits and sparse-grid techniques could result in savings in both memory and idling of computing nodes in a large-scale implementation. Finally, we mention that the Backward Monte Carlo scheme is likely to scale up favourably in a parallel computing environment. After all, the time stepping is explicit in nature and communication between different nodes could be optimized with proper domain decomposition mimicking the trajectories of the Hamiltonian orbits. 

\begin{acknowledgments}
The author is indebted to Guannan Zhang and Diego del-Castillo-Negrete for the enlightening discussions over the years regarding the Backward Monte Carlo methodology. This work was supported by the Academy of Finland grant no. 315278. Any subjective views or opinions expressed herein do not necessarily represent the views of the Academy of Finland or Aalto University.
\end{acknowledgments}

\bibliography{references}

\end{document}